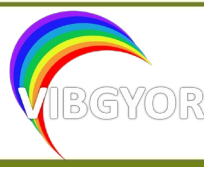

# A Novel Color Image Enhancement Method by the Transformation of Color Images to 2-D Grayscale Images

*Artyom M Grigoryan[1], Aparna John[1*] and Sos S Agaian[2]*

[1]Department of Electrical and Computer Engineering, The University of Texas at San Antonio, USA

[2]City University of New York, USA

**Abstract**

A novel method of color image enhancement is proposed, in which three or four color channels of the image are transformed to one channel 2-D grayscale image. This paper describes different models of such transformations in the RGB and other color models. Color image enhancement is achieved by enhancing first the transformed grayscale image and, then, transforming back the grayscale image into the colors. The color image enhancement is done on the transformed 2-D grayscale image rather than on the color image. New algorithms of color image enhancement are described in both frequency and time domains. The enhancement by this novel method shows good results. The enhancement of the image is measured with respect to the metric referred to as the Color Enhancement Measure Estimation (CEME).

## Introduction

A typical color image consists of three channels with each channel consisting of the image corresponding to the intensity values of particular color. In other words, color of the image is decomposed to color components and images corresponding to the color components are composed together to form a color image. Usually the image processing of the color image is done considering images in each channel as separate. In one of the most popular color image processing methods, the processing is done on each channel separately and the final enhanced image is composed by these enhanced channels. That is, image processing is done channel by channel. Another method of color image processing is based on the quaternion approach of color imaging [1-11]. In the quaternion approach, the color image is considered as the quaternion image and the channels in the color image considered as the components in the quaternion number [12-19]. In both these methods, processing is done on three channels of image, or on the 3-D images.

We are proposing a novel color image enhancement method in which the image processing is done on 2-D grayscale image only. In our method, the color image is transformed first to the 2-D grayscale image and image processing is done on 2-D grayscale image. This paper describes transformation of color image to 2-D grayscale image by a few proposed transformation models like the $2 \times 2$, $2 \times 3$, column and row transformation models. The image enhancement effects by frequency domain enhancement method, such as the alpha-rooting, and spatial domain enhancement method by the histogram equalization by the novel method is studied.

## Transformation Models

In the novel method of color image enhancement, color images are transformed to 2-D grayscale images and the enhancement algorithms, which work well in 2-D grayscale images, can be used in the newly transformed 2-D grayscale-color image [20]. There are many different models which can be considered for this transformation. A few of these transformation models are the 2 × 2 model, 2 × 3 model, column model and row model. First, we describe the 2 × 2 model of color-to-gray transformation for RGB color image. There are three channels of RGB image R, G, and B, and the luminous of this RGB image is obtained by the relation

$$I = 0.3R + 0.59G + 0.11B. \quad (1)$$

**\*Corresponding author:** *Aparna John, Department of Electrical and Computer Engineering, The University of Texas at San Antonio, One UTSA Circle, San Antonio, TX-78249, USA, E-mail: aparna.john@utsa.edu*

In the 2 × 2 model, the new transformed 2-D grayscale image is obtained by arranging side by side the pixel values of I, R, G, and B of each pixel. For instance the table below depicts one of the ways the 2 × 2 model is obtained.

| | | |
|---|---|---|
| I(n,m) | R(n,m) | … |
| G(n,m) | B(n,m) | … |
| … | … | … |

Therefore, the image of size M × N × 3 is transformed to (2M) × (2N) 2-D grayscale image.

| I(0,0) | R(0,0) | I(0,1) | R(0,1) | I(0,2) | R(0,2) | … |
|---|---|---|---|---|---|---|
| G(0,0) | B(0,0) | G(0,1) | B(0,1) | G(0,2) | B(0,2) | … |
| I(1,0) | R(1,0) | I(1,1) | R(1,1) | I(1,2) | R(1,2) | … |
| G(1,0) | B(1,0) | G(1,1) | B(1,1) | G(1,2) | B(1,2) | … |
| I(2,0) | R(2,0) | I(2,1) | R(2,1) | I(2,2) | R(2,2) | … |
| G(2,0) | B(2,0) | G(2,1) | B(2,1) | G(2,2) | B(2,2) | … |
| … | … | … | … | … | … | … |

The 2 × 3 model is another transformation model, in which the R, G, and B are arranged in such a way that a single unit so formed is a 2 × 3 in size. There are many different ways to arrange R, G, and B pixel values. One of the many different possibilities is as depicted below in the table.

| R(0,0) | G(0,0) | B(0,1) | R(0,2) | G(0,2) | B(0,3) | R(0,4) | G(0,4) | B(0,5) | … |
|---|---|---|---|---|---|---|---|---|---|
| B(0,0) | R(0,1) | G(0,1) | B(0,2) | R(0,3) | G(0,3) | B(0,4) | R(0,5) | G(0,5) | … |
| R(1,0) | G(1,0) | B(1,1) | R(1,2) | G(1,2) | B(1,3) | R(1,4) | G(1,4) | B(1,5) | … |
| B(1,0) | R(1,1) | G(1,1) | B(1,2) | R(1,3) | G(1,3) | B(1,4) | R(1,5) | G(1,5) | … |
| R(2,0) | G(2,0) | B(2,1) | R(2,2) | G(2,2) | B(2,3) | R(2,4) | G(2,4) | B(2,5) | … |
| B(2,0) | R(2,1) | G(2,1) | B(2,2) | R(2,3) | G(2,3) | B(2,4) | R(2,5) | G(2,5) | … |
| … | … | … | … | … | … | … | … | … | … |

The color image of size $M \times N \times 3$ is transformed to a 2-D grayscale image of size $(2M) \times (3N/2)$.

| I(0,0) | I(0,1) | I(0,2) | … |
|---|---|---|---|
| R(0,0) | R(0,1) | R(0,2) | … |
| G(0,0) | G(0,1) | G(0,2) | … |
| B(0,0) | B(0,1) | B(0,2) | … |
| I(1,0) | I(1,1) | I(1,2) | … |
| R(1,0) | R(1,1) | R(1,2) | … |
| G(1,0) | G(1,1) | G(1,2) | … |
| B(1,0) | B(1,1) | B(1,2) | … |
| … | … | … | … |

The table below depicts the row model and in a row model the color image of size M × N × 3 together with the luminous component I is transformed to a 2-D gray-scale image of size (4M) × N. We can also opt out the luminous component in the new model. Then the new transformed image is of size (3M) × N.

And the column model of transformation is as depicted in table below. The color image of size M × N × 3 together with the luminous component I is transformed to an image of size M × (4N). It is optional to include the luminous component I in the new transformed model.

| I(0,0) | R(0,0) | G(0,0) | B(0,0) | I(0,1) | R(0,1) | G(0,1) | B(0,1) | … |
|---|---|---|---|---|---|---|---|---|
| I(1,0) | R(1,0) | G(1,0) | B(1,0) | I(1,1) | R(1,1) | G(1,1) | B(1,1) | … |
| I(2,0) | R(2,0) | G(2,0) | B(2,0) | I(2,1) | R(2,1) | G(2,1) | B(2,1) | … |
| … | … | … | … | … | … | … | … | … |

## Enhancement Methods

### Alpha-rooting method

In the alpha-rooting method of image enhancement [14,16,21-24], for each frequency point (p,s), the magnitude of the discrete transform is modified as

$$\left|F_{p,s}\right| \to \left|F_{p,s}\right|^{\alpha} . \quad (2)$$

### Histogram equalization of color images

The histogram is an important concept in imaging, which gives the number (cardinality) of pixels of the image with the given level of intensity r. For the image $f_{n,m}$ of size M × N, the histogram is a non-negative function,

$$h(r) = card \; \{(m, n); f_{m, n} = r, m = 0,1,\dots, (M-1), n = 0,1,\dots, (N-1)\}. \quad (3)$$

The histogram is normalized, h(r) = h(r)/(MN) so that 0 < h(r) < 1. In the method of histogram equalization, a monotonic increasing grayscale transformation w is used to straighten the curve of the histogram.

$$w: r = f_{m,n} \to s = w(r) \to g_{m,n} . \quad (4)$$

It is assumed that the value of the image as random number r from the interval $[f_{min}, f_{max}]$ is transformed to the random value s which is approximately uniformly distributed in the interval $[w_{min}, w_{max}] = [w(f_{min}), w(f_{max})]$. This monotonic transformation w is calculated as

$$w(r) = w_{min} + [w_{max} - w_{min}]F(r). \quad (5)$$

Here, F(r) is the distribution function of the intensity and is calculated as

$$F(r) = \sum_{k=f_{min}}^{r} h(k), \; Where \; r = [f_{min}, f_{max}] \quad (6)$$

and F(r) = 0, when $r < f_{min}$. In the case when $[w_{min}, w_{max}] = [0, 255]$, the histogram equalization is calculated by the transformation

$$r \to s \left\{ \left[ 255 \sum_{k=0}^{r} h(k) \right] \; for \; r = 0,1, \dots, 255 \right. \quad (7)$$

### Enhancement Measure Estimation (EME)

The Enhancement Measure Estimation (EME) for grayscale images, is an enhancement measure [25-28]

based on the contrast of the images in one channel 2-D image or grayscale image. The proposed enhancement measure is a modification of Weber's law which basically explains that the visual perception of the contrast is in-dependent of luminance and low spatial frequency. The metric EME relates to the Weber's law that states that the perceived change in stimulus proportional to initial stimuli, and the Fechner's law, which states that the perception and stimulus are logarithmically related. That is, the visually perceived intensity value is proportional to the logarithm of the actual intensity. To calculate the EME value, the 2-D discrete image of size N × M is divided by $k_1 k_2$ blocks of size $L_1 \times L_2$ blocks each, where $k_n = \lfloor N_n/L_n \rfloor$, for n = 1, 2. When an image is enhanced,

$$f \to \hat{f}. \quad (8)$$

Here $f$ and $\hat{f}$, referring to the original and en-hanced image respectively, the EME value of enhanced image is calculated by

$$EME(\hat{f}) = \frac{1}{k_1 k_2} \sum_{k=1}^{k_1} \sum_{l=1}^{k_2} 20 \log_{10}\left[\frac{\max_{k,l}(\hat{f})}{\min_{k,l}(\hat{f})}\right] \quad (9)$$

The EME of the original image is obtained by replacing $\hat{f}$ in Eq. 9 by $f$

**Color Enhancement Measure Estimation (CEME)**

The Color Enhancement Measure Estimation (CEME), is an enhancement measure [25-28] equivalent of EME measure but CEME metric is the measure of visual perception of color images. To calculate the CEME value, the 2-D discrete image of size N × M is divided [12] by $k_1 k_2$ blocks of size $L_1 \times L_2$ blocks each, where $k_n = \lfloor N_n/L_n \rfloor$ for n = 1, 2. When the original image $f$ is enhanced to $\hat{f}$,

$$f = (f_R, f_G, f_B) \to \hat{f} = (\hat{f}_R, \hat{f}_G, \hat{f}_B), \quad (10)$$

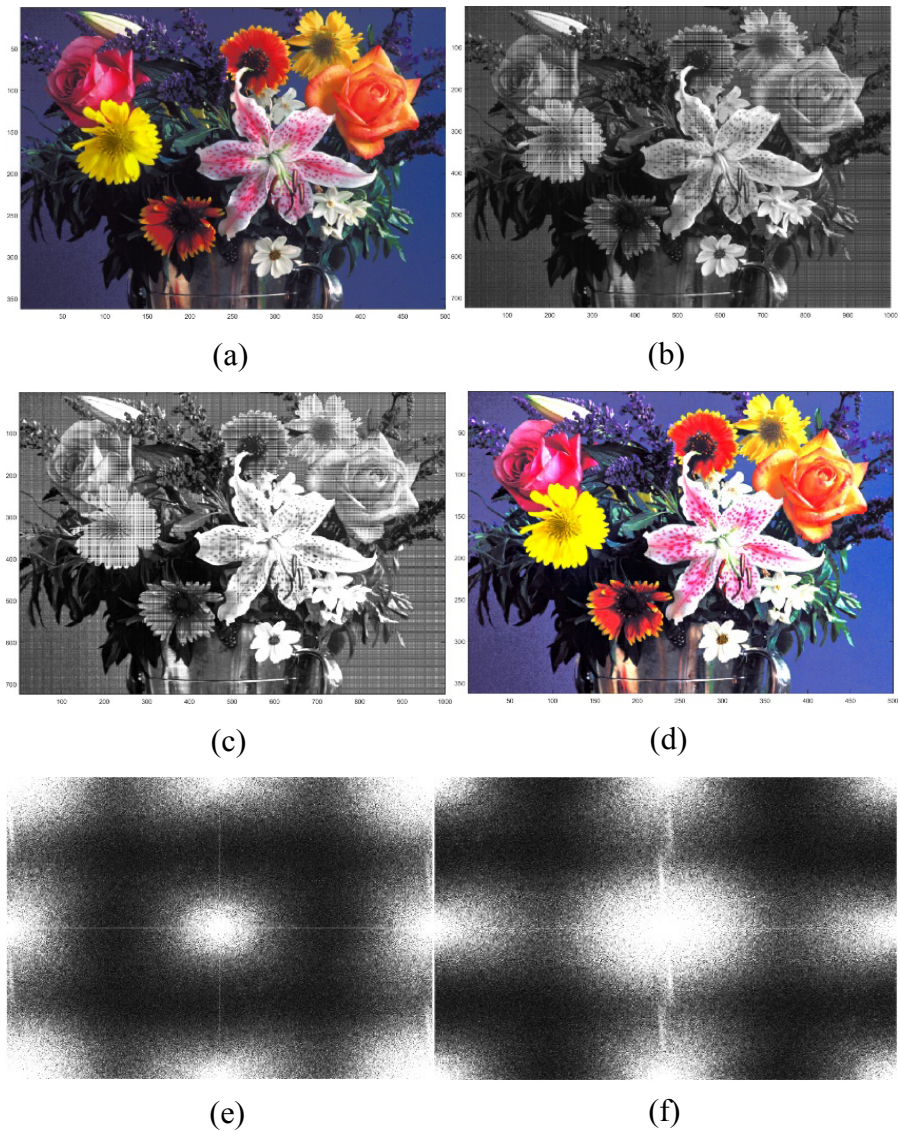

**Figure 1:** a) Original image "flower.tif"; b) 2-D grayscale image of (a) by 2 × 2 transformation model; c) Alpha-rooting of 2-D grayscale image in (b); d) RGB color image after converting back to color the image in (c); e) 2-D DFT of image in (b); f) Center-shifted 2-D DFT in (e).

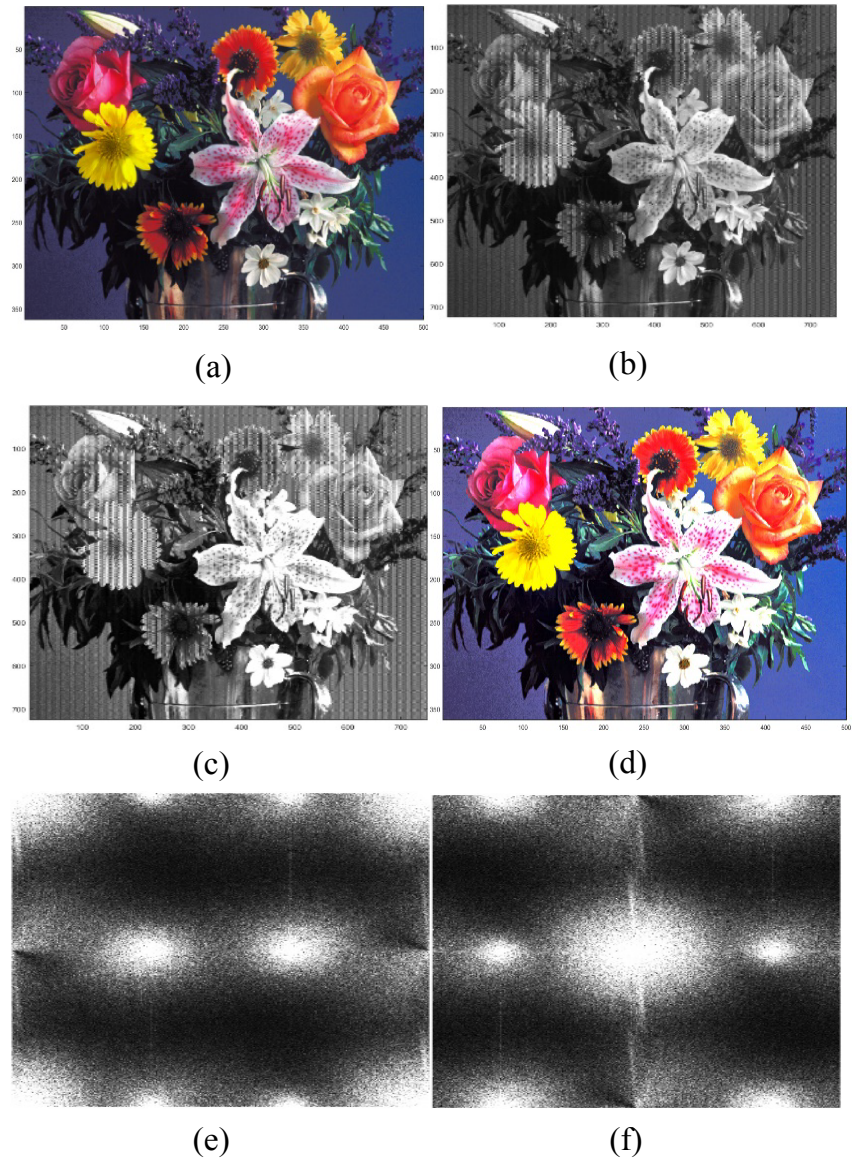

**Figure 2:** a) Original Image "flower.tif"; b) 2-D grayscale image of (a) by 2 × 3 Transformation model; c) Alpha-Rooting of 2-D grayscale image in (b); d) RGB color image after converting back to color the image in (c); e) 2-D DFT of image in (b); f) Center-shifted 2-D DFT in (e).

**Table 1:** Alpha, EME and CEME values of alpha-rooting method on "flower.tif".

| Image "flower.tif" | EME | CEME |
|---|---|---|
| Original Image | - | 28.7506 |
| 2-D Grayscale Image (2 × 2 model) | 22.1548 | - |
| 2-D Grayscale Image (2 × 2 model) - After Alpha-Rooting (alpha = 0.97) | 29.0065 | - |
| RGB Color image - After Alpha-Rooting (alpha = 0.97) of 2-D Grayscale Image (2 × 2 model) | - | 36.9647 |
| | | |
| 2-D Grayscale Image (2 × 3 model) | 23.6574 | |
| 2-D Grayscale Image (2 × 3 model) - After Alpha-Rooting (alpha = 0.97) | 30.9713 | |
| RGB Color image - After Alpha-Rooting (alpha = 0.97) of 2-D Grayscale Image (2 × 3 model) | | 36.4484 |
| | | |
| 2-D Grayscale Image (column model) | 23.0670 | |
| 2-D Grayscale Image (column model) - After Alpha-Rooting (alpha = 0.97) | 30.2462 | |
| RGB Color image - After Alpha-Rooting (alpha = 0.97) of 2-D Grayscale Image (column model) | | 37.0227 |
| | | |
| 2-D Grayscale Image (row model) | 23.4101 | |
| 2-D Grayscale Image (row model) - After Alpha-Rooting (alpha = 0.97) | 30.6765 | |
| RGB Color image - After Alpha-Rooting (alpha = 0.97) of 2-D Grayscale Image (row model) | | 36.9149 |

The CEME value of the enhanced color image is calculated by

$$CEME(\hat{f}) = \frac{1}{k_1 k_2} \sum_{k=1}^{k_1} \sum_{l=1}^{k_2} 20 \log_{10} \left[ \frac{max_{k,l}(\hat{f}_R, \hat{f}_G, \hat{f}_B)}{min_{k,l}(\hat{f}_R, \hat{f}_G, \hat{f}_B)} \right] \quad (11)$$

Here $f_R, f_G, f_B$ and $\hat{f}_R, \hat{f}_G, \hat{f}_B$, refer respectively to R, G, and B channel of the original and enhanced image. In our experiment, $\hat{f}_R, \hat{f}_G, \hat{f}_B$ are enhanced color channels which are obtained by reconverting back the enhanced 2-D grayscale image. The CEME of the original image is similar to Eq. 11 but $\hat{f}$ replaced by $f$,

$$CEME(f) = \frac{1}{k_1 k_2} \sum_{k=1}^{k_1} \sum_{l=1}^{k_2} 20 \log_{10} \left[ \frac{max_{k,l}(f_R, f_G, f_B)}{min_{k,l}(f_R, f_G, f_B)} \right] \quad (12)$$

## Experimental Results

A few image results of the enhancement by the novel method are shown below. The enhancement in the frequency domain by the alpha-rooting method and the enhancement in the spatial domain by the histogram equalization show good enhancement of images. Figure 1 shows the original image "flower.tif" and the transformation of the color image to 2-D grayscale image by the 2 × 2 transformation model (Figure 1a and Figure 1b). The alpha-rooting method is applied to the transformed 2-D grayscale image and then converted back to the color image (Figure 1c and Figure 1d). The optimum value of alpha chosen for the enhancement is the alpha that gives a maximum CEME value for the enhanced color image. For "flower.tif" image, the alpha value is 0.97. The EME value of the 2-D grayscale image after transforming by 2 × 2 transformation model and after performing alpha-rooting method is tabulated in Table 1. The CEME value of the image after converting back to the color image is higher than the CEME value of the original image. Figure 1e and Figure 1f shows the 2-D DFT and the center-shifter 2-D DFT of the image in Figure 1b, respectively. The periodicity in the row and column in 2-D DFT is seen based on the 2 × 2 model.

Figure 2 shows the results of the transformation of the color image by the 2 × 3 transformation model. Figure 2a and Figure 2b show the original image and its transformation to 2-D grayscale image by the 2 × 3 model. Figure 2c shows the image after performing the alpha-rooting method of enhancement. The optimum choice of alpha in this case is also 0.97 and at alpha equals 0.97 the CEME value of the color image is maximum. Figure 2d shows the color image after converting back the 2-D grayscale image processed by alpha-rooting method. Figure 2e and Figure 2f shows the 2-D DFT and the center-shifted 2-D DFT of the transformed grayscale image shown in part 2b.

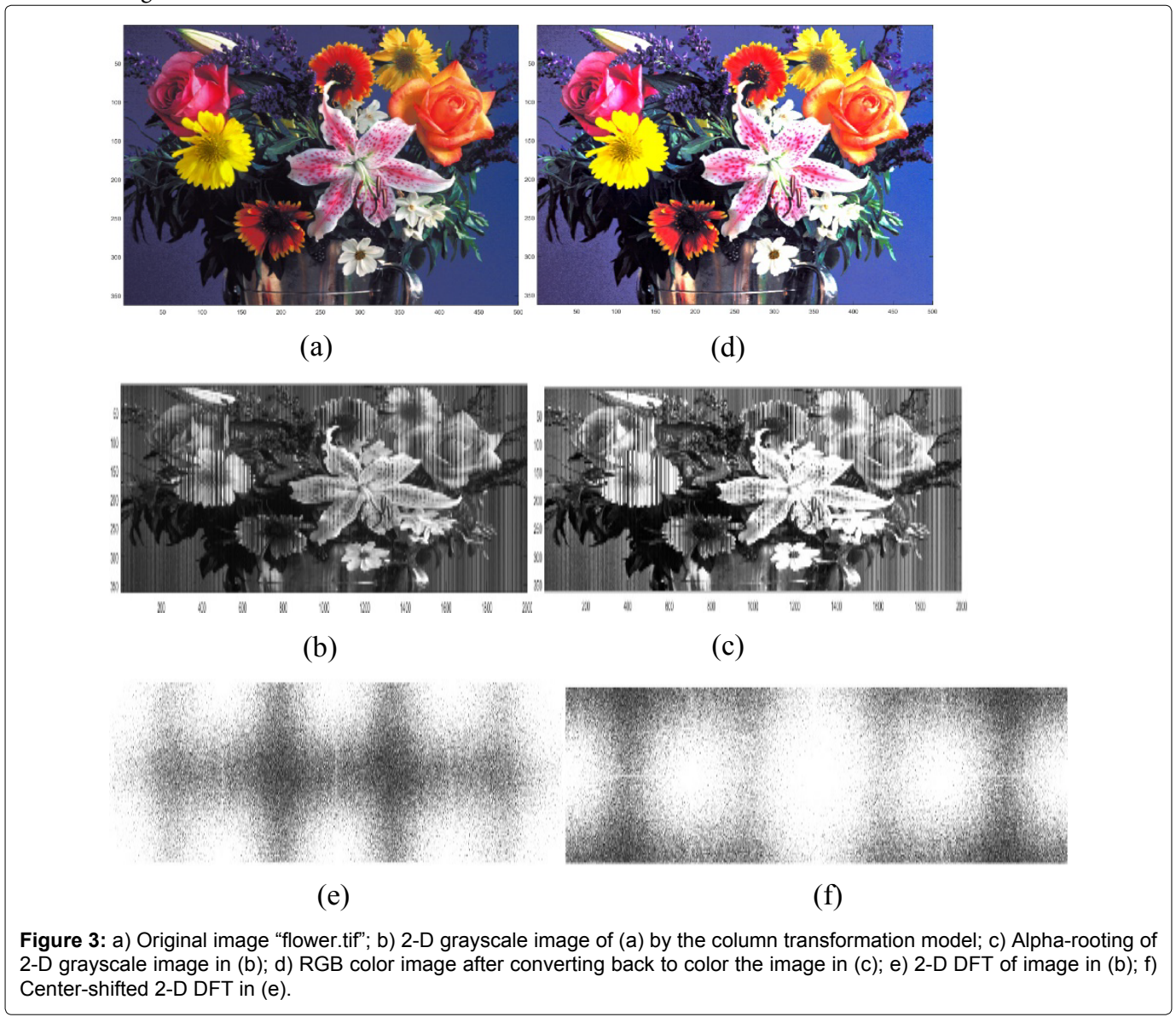

**Figure 3:** a) Original image "flower.tif"; b) 2-D grayscale image of (a) by the column transformation model; c) Alpha-rooting of 2-D grayscale image in (b); d) RGB color image after converting back to color the image in (c); e) 2-D DFT of image in (b); f) Center-shifted 2-D DFT in (e).

Figure 3 shows the results of alpha-rooting method of color image when transformed by the column model. The alpha value chosen in this method is 0.97. The 2-D DFT and center-shifted 2-D DFT show a periodicity in the column. Figure 4 shows image results after the color image is transformed by the row model. The alpha values chosen in alpha-rooting method in each transformation model and the respective EME and CEME values of the images are tabulated in Table 1.

In Table 1, we see that the best alpha value for alpha-rooting of "flower.tif" is 0.97, irrespective of the transformation model used for transforming the color image to the 2-D grayscale image. The CEME value of the enhanced color image is higher than the CEME value of the original image, which is 28.7506. The CEME value of the enhanced color image is almost the same for all four different models of transformation. The EME value of the 2-D grayscale image of the enhanced image is also higher than the EME of the transformed 2-D grayscale image of the original image. EME values also show almost nearby values for all transformation models.

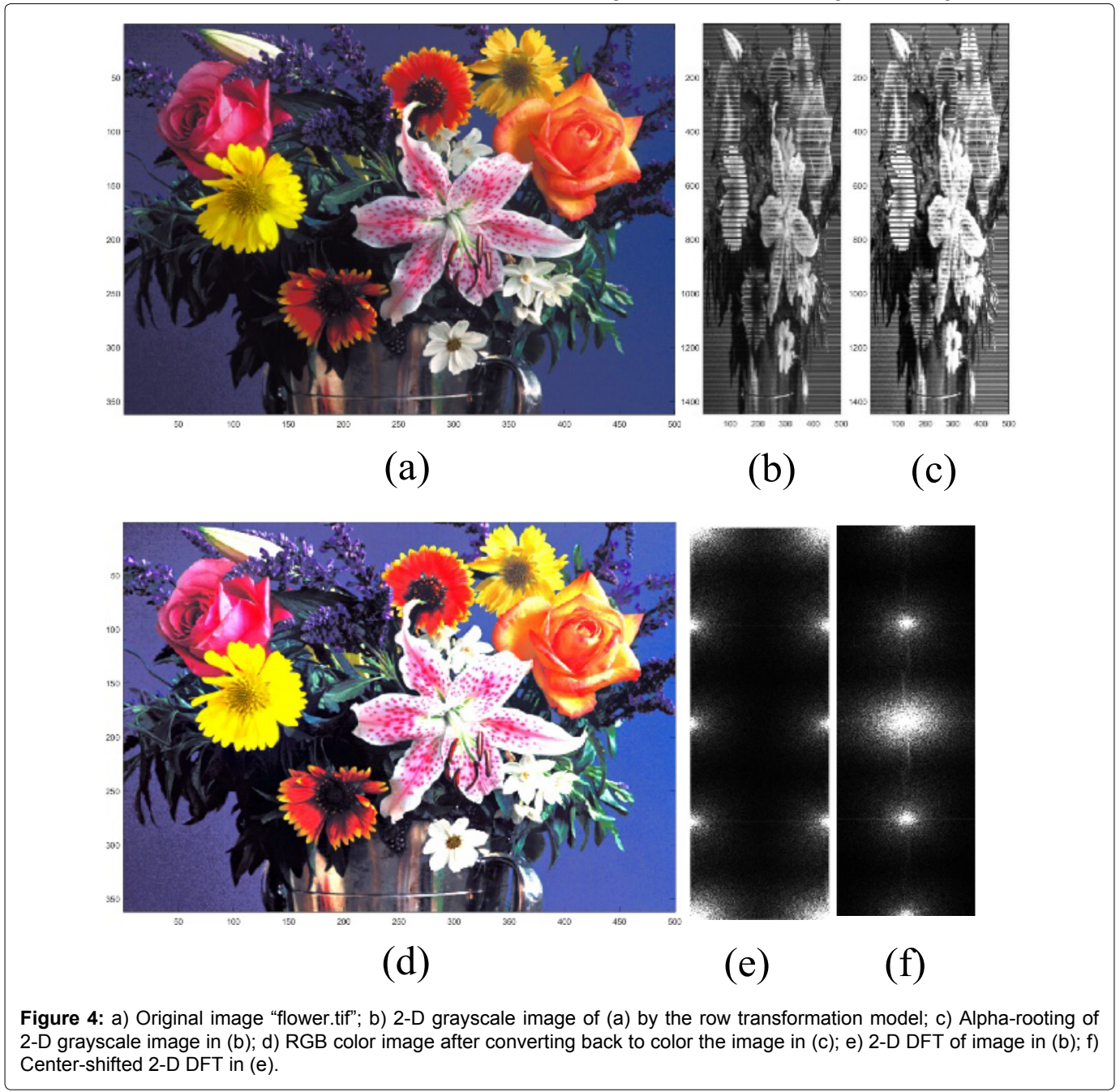

**Figure 4:** a) Original image "flower.tif"; b) 2-D grayscale image of (a) by the row transformation model; c) Alpha-rooting of 2-D grayscale image in (b); d) RGB color image after converting back to color the image in (c); e) 2-D DFT of image in (b); f) Center-shifted 2-D DFT in (e).

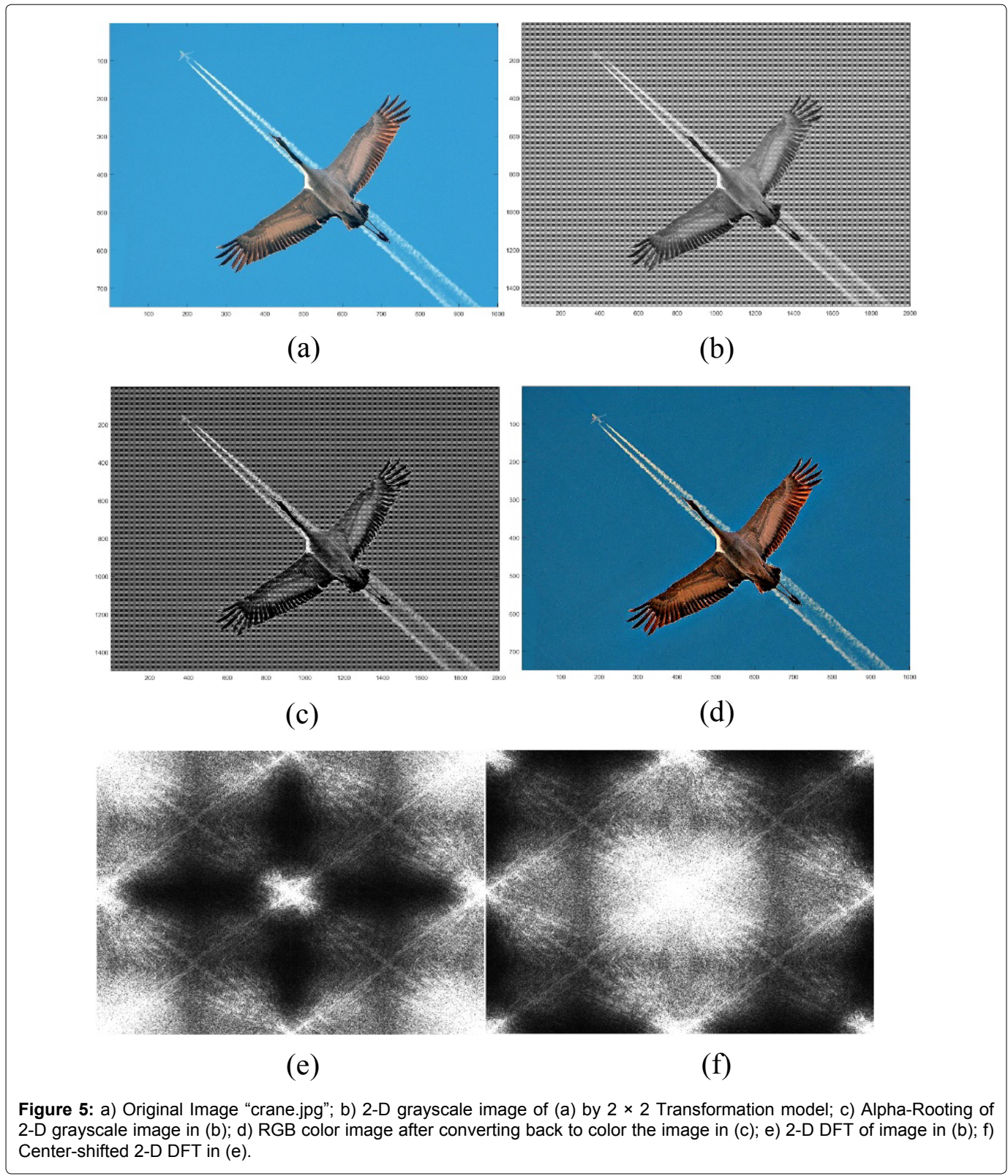

**Figure 5:** a) Original Image "crane.jpg"; b) 2-D grayscale image of (a) by 2 × 2 Transformation model; c) Alpha-Rooting of 2-D grayscale image in (b); d) RGB color image after converting back to color the image in (c); e) 2-D DFT of image in (b); f) Center-shifted 2-D DFT in (e).

Figure 5 is the image results of the alpha-rooting method of the original image "crane.jpg" (Figure 5a) after transforming the color image to 2-D grayscale im-age by 2 × 2 model. (Figure 5b and Figure 5c) shows respectively the 2-D grayscale image of the original image and after enhancing by alpha-rooting method with alpha equals 0.88. When alpha equals 0.88, the CEME value is maximum for the color image obtained after converting back the enhanced 2-D grayscale image. Figure 5d shows the final enhanced color image. Figure 5e and Figure 5f shows the 2-D DFT and the center-shifted 2-D DFT of the 2-D grayscale image in Figure 5b. Since the transformation model chosen is 2 × 2, the 2-D DFT shows a periodicity in the row and column direction similar to the spatial domain transformation of the 2-D grayscale image.

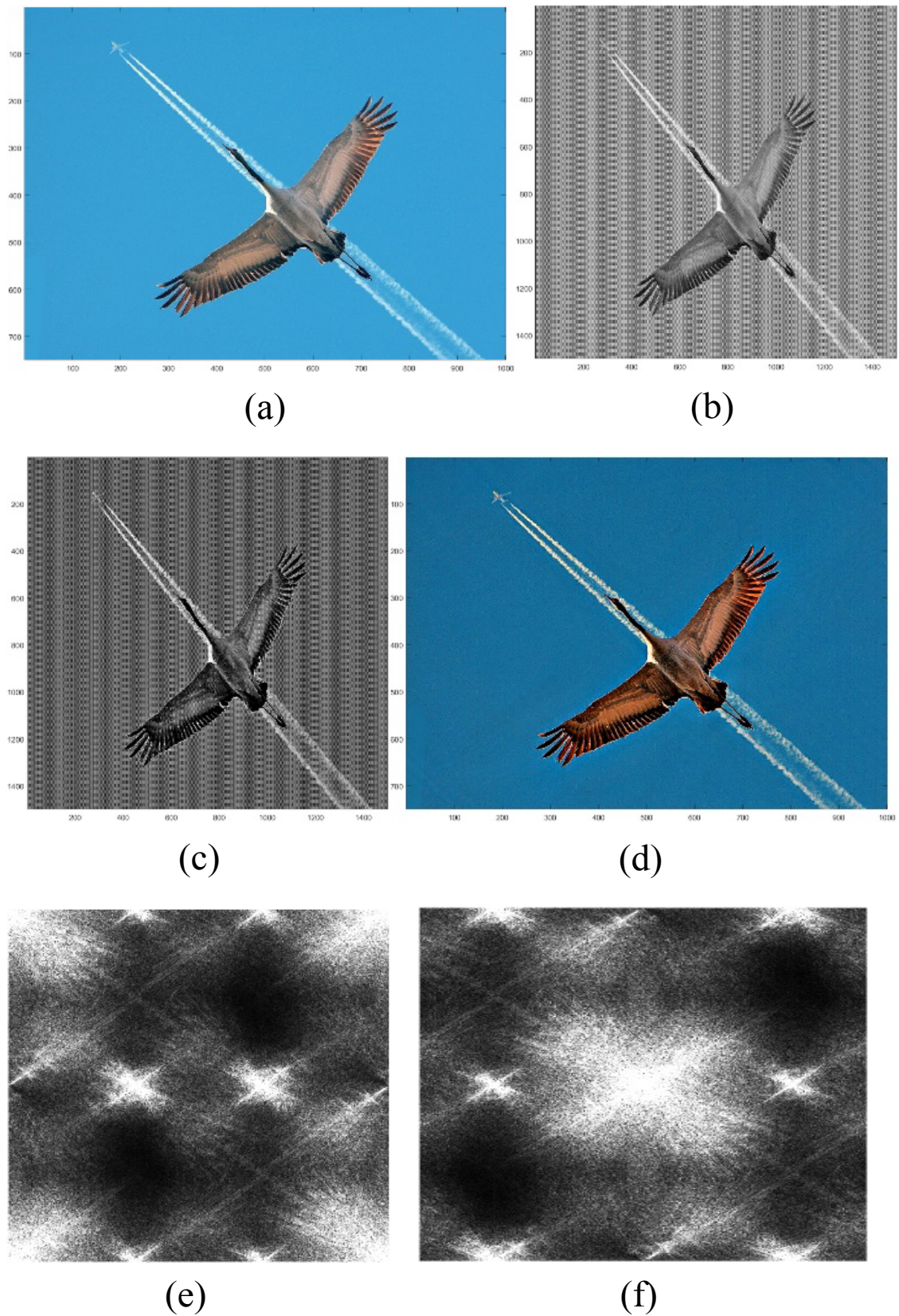

**Figure 6:** a) Original Image "crane.jpg"; b) 2-D grayscale image of (a) by 2 × 3 Transformation model; c) Alpha-Rooting of 2-D grayscale image in (b); d) RGB color image after converting back to color the image in (c); e) 2-D DFT of image in (b); f) Center-shifted 2-D DFT in (e).

Figure 6 shows the image results of the original image "crane.jpg" after transforming the color image by 2 × 3 model. Figure 6a shows the 2-D grayscale image after transforming by the 2 × 3 model. When the transformed image in Figure 6b is enhanced by the alpha-rooting method, the enhancement of the image is as shown in Figure 6c. The optimum alpha value for the alpha-rooting method is 0.88. Figure 6d shows the color image after converting back the enhanced 2-D grayscale image in Figure 6c. When alpha equals 0.88 the color image in Figure 6d gives the maximum CEME value. Figure 6e and Figure 6f shows respectively the 2-D DFT and centershifted 2-D DFT of the image in Figure 6b. The Fourier transform shows a periodicity in the 2 × 3 model in the row and column direction.

In Figure 7, the original image "crane.jpg" (Figure 7a) is transformed by the column model to the correspond-ing 2-D grayscale image. Figure 7b shows that the image is four times wider in the column direction as compared to the column width of the original image while the row size of the transformed image remains same as the original image. The transformed 2-D grayscale image in Figure 7b is enhanced by the alpha-rooting method with the best alpha which is 0.88 and the enhanced image is shown in Figure 7c. The enhanced color image is shown in Figure 7d. The 2-D DFT and the center-shifted 2-D DFT in (Figure 7e and Figure 7f) show a periodicity in the column direction.

Figure 8 shows enhancement results after transforming the color image (Figure 8a) to 2-D grayscale image by row method. The row size of the transformed grayscale image (Figure 8b) is four times the row size of the original image while the column size is same for both original image and transformed grayscale image. Figure 8c is the enhanced 2-D grayscale image after performing alpha-rooting method with alpha equals 0.88. The 2-D DFT and the center-shifted 2-D DFT given in Figure 8e and Figure 8f show a periodicity in the row direction.

In Table 2, one can see that the alpha value chosen for the alpha-rooting method is the same for all transformation models. The CEME of the color image after enhancing by the novel transformation method is higher than the CEME value of the original image (CEME = 28.7627). The CEME value of the obtained by all transformation models has values close to 46. The EME value of the 2-D grayscale transformed image has value 27 and EME of the enhanced grayscale image by the alpha-rooting method has value in the range 42-44 for all models.

The figures that follow show the image results by the spatial transformation by the histogram equalization after transforming the color image by different transformation models. In Figure 9, the original image "flower.tif" (Figure 9a) is first transformed to the 2-D grayscale im-age by the 2 × 2 model (Figure 9b) and the histogram equalization is calculated on the 2-D grayscale image; the result is shown in

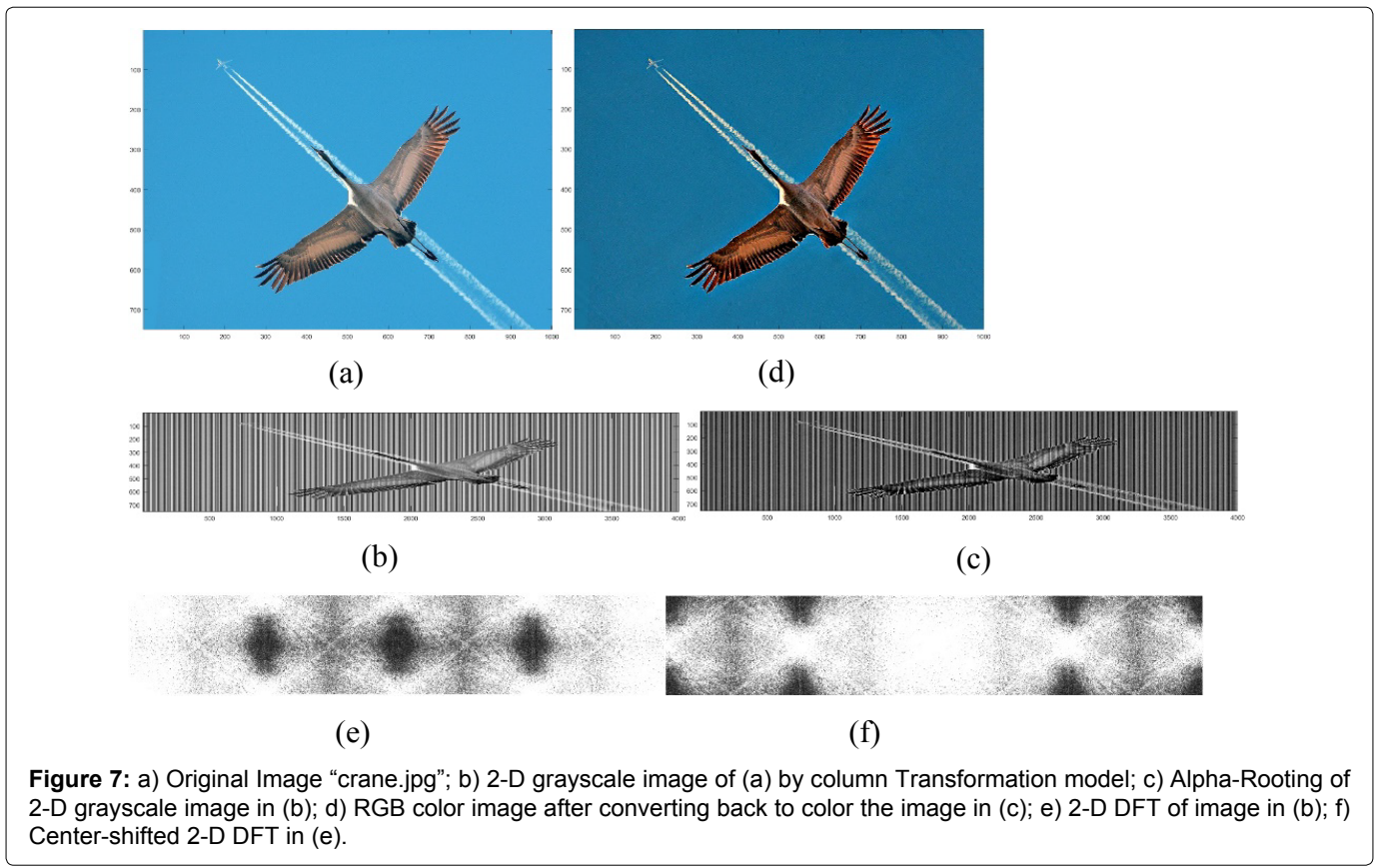

**Figure 7:** a) Original Image "crane.jpg"; b) 2-D grayscale image of (a) by column Transformation model; c) Alpha-Rooting of 2-D grayscale image in (b); d) RGB color image after converting back to color the image in (c); e) 2-D DFT of image in (b); f) Center-shifted 2-D DFT in (e).

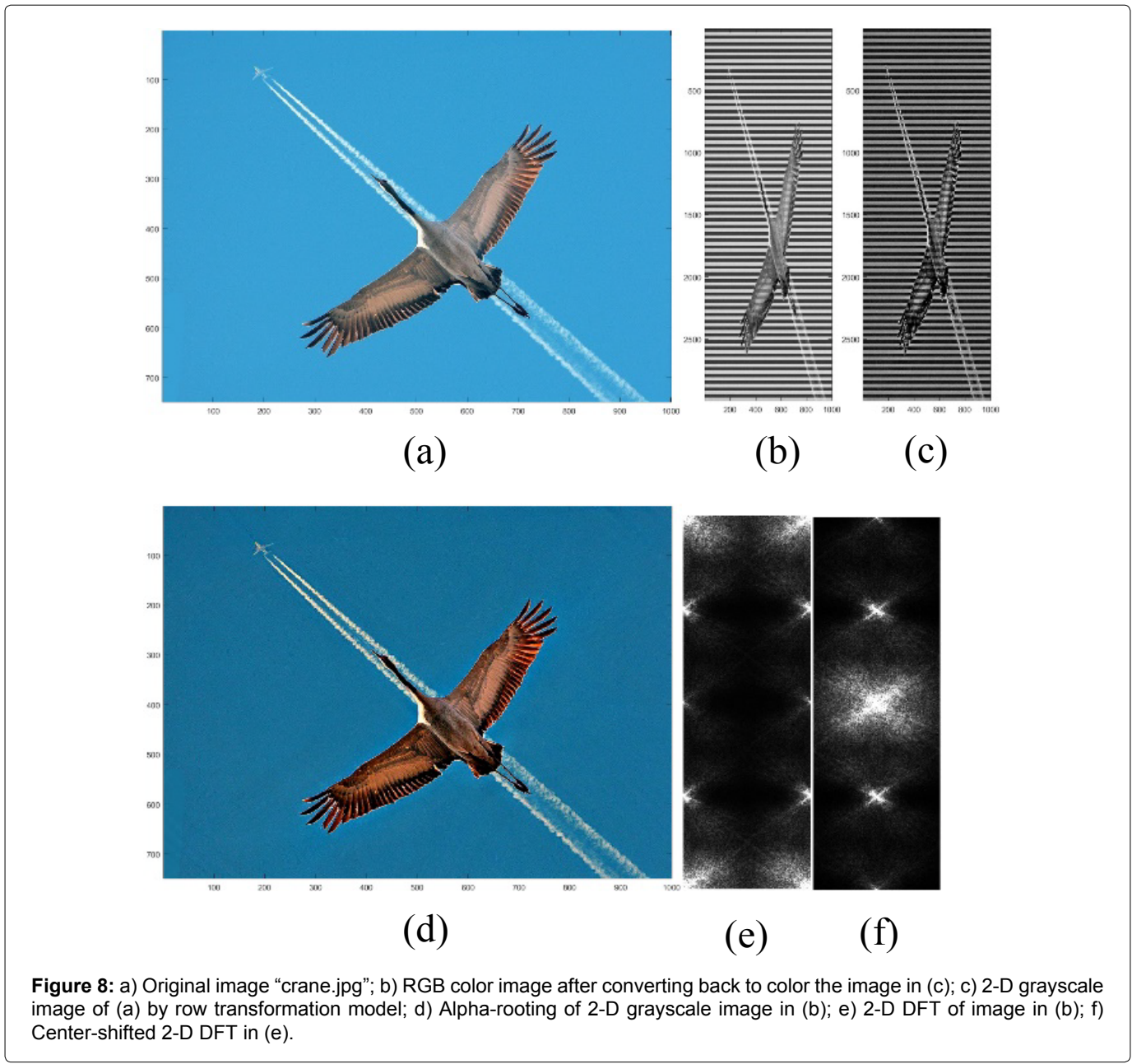

**Figure 8:** a) Original image "crane.jpg"; b) RGB color image after converting back to color the image in (c); c) 2-D grayscale image of (a) by row transformation model; d) Alpha-rooting of 2-D grayscale image in (b); e) 2-D DFT of image in (b); f) Center-shifted 2-D DFT in (e).

**Table 2:** Alpha, EME and CEME values of alpha-rooting method on "crane.jpg".

| Image | EME | CEME |
|---|---|---|
| Original Image | - | 28.7627 |
| 2-D Grayscale Image (2 × 2 model) | 27.4693 | - |
| 2-D Grayscale Image (2 × 2 model) - After Alpha-Rooting (alpha = 0.88) | 43.0167 | - |
| RGB Color image - After Alpha-Rooting (alpha = 0.88) of 2-D Grayscale Image (2 × 2 model) | - | 46.2691 |
|  |  |  |
| 2-D Grayscale Image (2 × 3 model) | 27.8003 |  |
| 2-D Grayscale Image (2 × 3 model) - After Alpha-Rooting (alpha = 0.88) | 44.3429 |  |
| RGB Color image - After Alpha-Rooting (alpha = 0.88) of 2-D Grayscale Image (2 × 3 model) |  | 46.5222 |
|  |  |  |
| 2-D Grayscale Image (row model) | 27.6265 |  |
| 2-D Grayscale Image (row model) - After Alpha-Rooting (alpha = 0.88) | 43.0678 |  |
| RGB Color image - After Alpha-Rooting (alpha = 0.88) of 2-D Grayscale Image (row model) |  | 46.3382 |
|  |  |  |
| 2-D Grayscale Image (column model) | 27.6164 |  |
| 2-D Grayscale Image (column model) - After Alpha-Rooting (alpha = 0.88) | 42.9896 |  |
| RGB Color image - After Alpha-Rooting (alpha = 0.88) of 2-D Grayscale Image (column model) |  | 46.3066 |

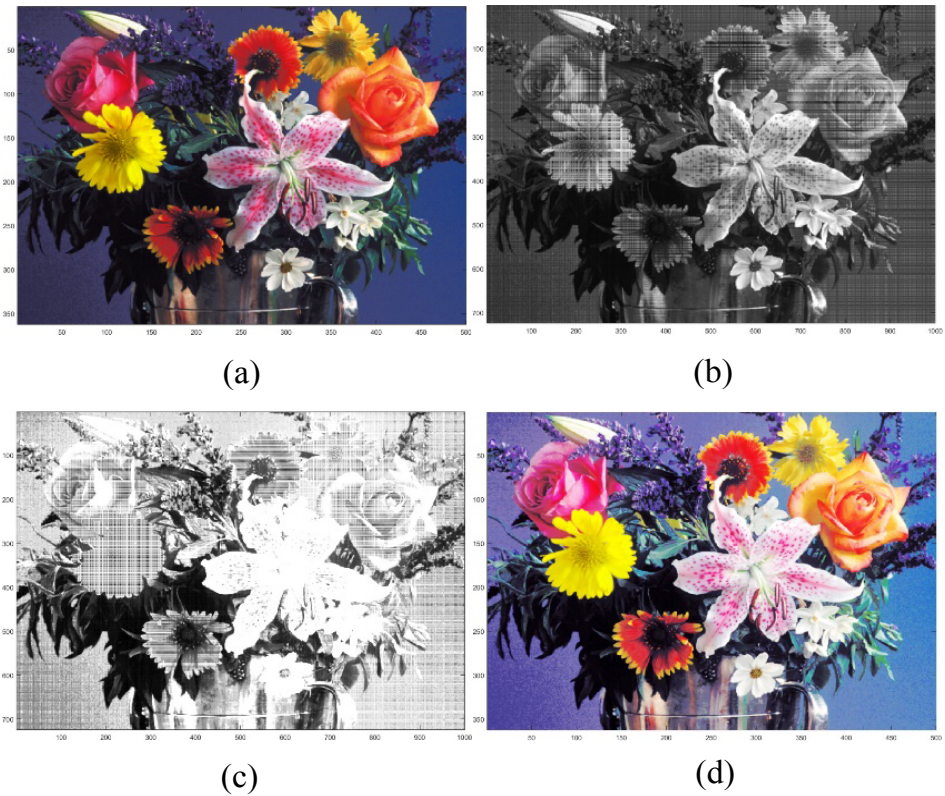

**Figure 9:** a) Original image "flower.tif"; b) 2-D Grayscale image transformed by 2 × 2 model; c) Histogram equalized 2-D grayscale image in (b); d) RGB color image after converting the 2-D grayscale image in (c) to color image.

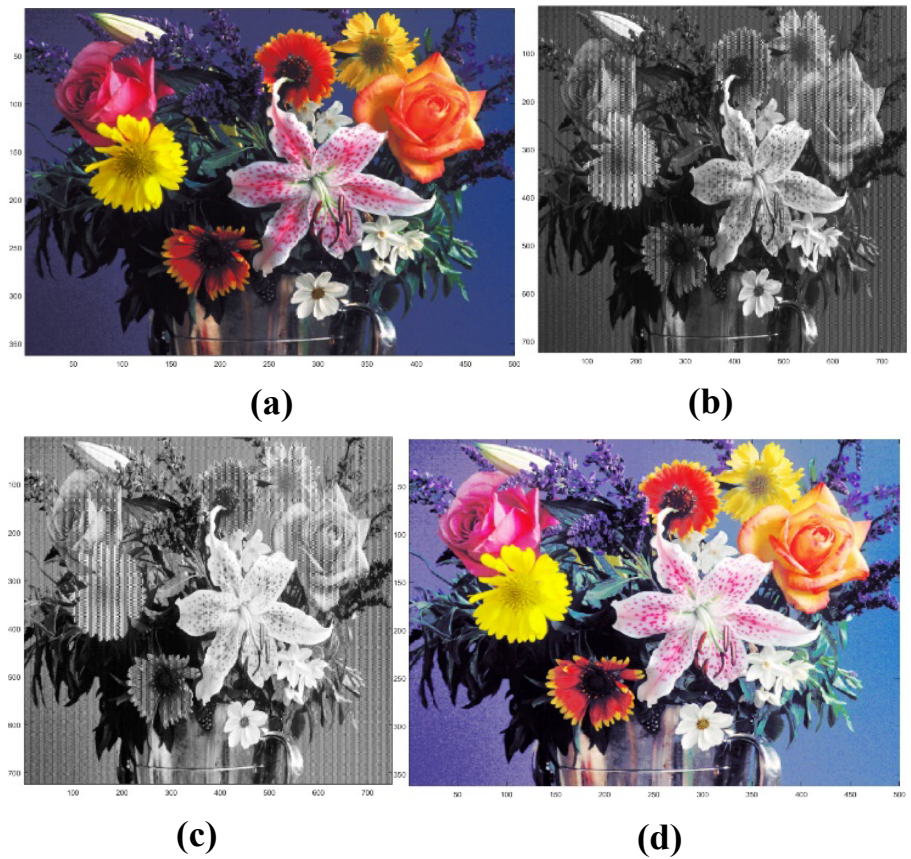

**Figure 10:** a) Original Image "flower.tif"; b) 2-D Grayscale image transformed by 2 × 3 model; c) Histogram equalized 2-D grayscale image in (b); d) RGB color image after converting the 2-D grayscale image in (c) to color image.

Figure 9c. Figure 9d shows the final color image after converting back the histogram equalized grayscale image in Figure 9c.

Figure 10 is the enhancement result after the color image is transformed to 2 × 3 model 2-D grayscale image. Figure 10b and Figure 10c are respectively the 2 × 2 model 2-D grayscale image of the original image in Figure 10a, and the histogram equalized image in Fig-ure 10b. After enhancing the transformed 2-D grayscale image by the histogram equalization, it is converted back to color image, the result is shown in Figure 10d.

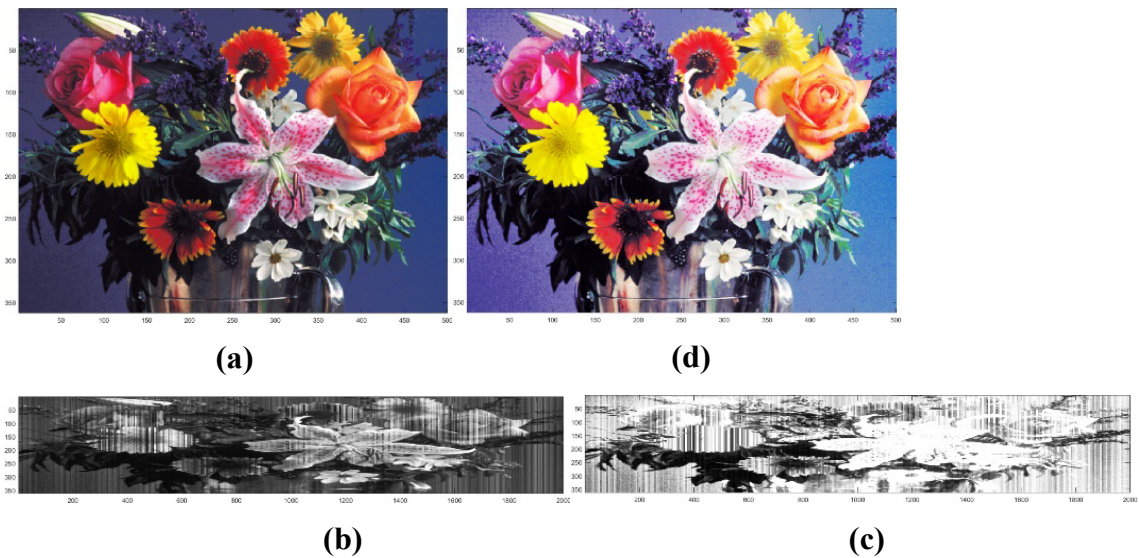

**Figure 11:** a) Original Image "flower.tif"; b) 2-D Grayscale image transformed by column model; c) Histogram equalized 2-D grayscale image in (b); d) RGB color image after converting the 2-D grayscale image in (c) to color image.

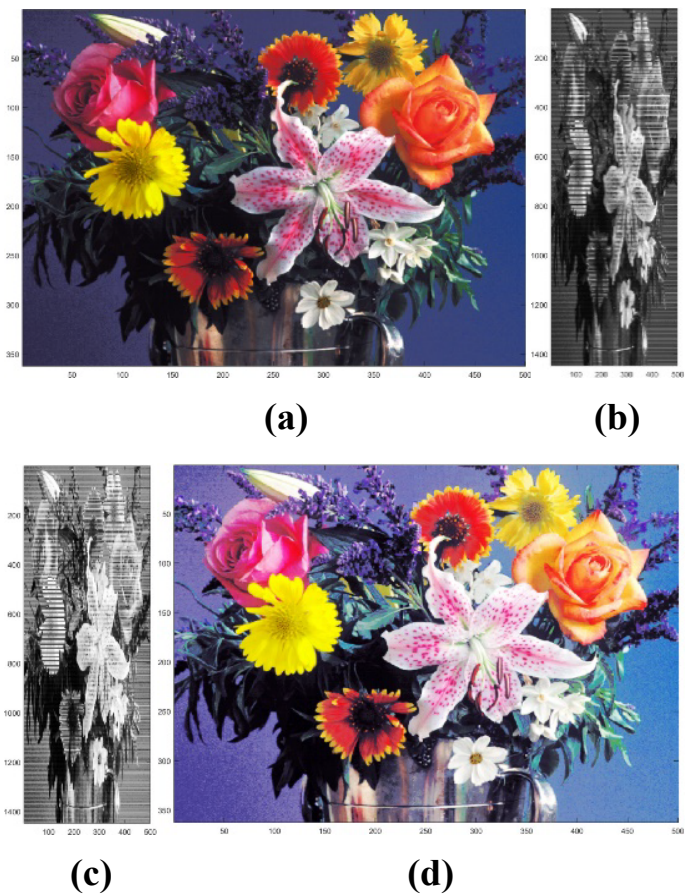

**Figure 12:** a) Original Image "flower.tif"; b) 2-D Grayscale image transformed by row model; c) Histogram equalized 2-D grayscale image in (b); d) RGB color image after converting the 2-D grayscale image in (c) to color image.

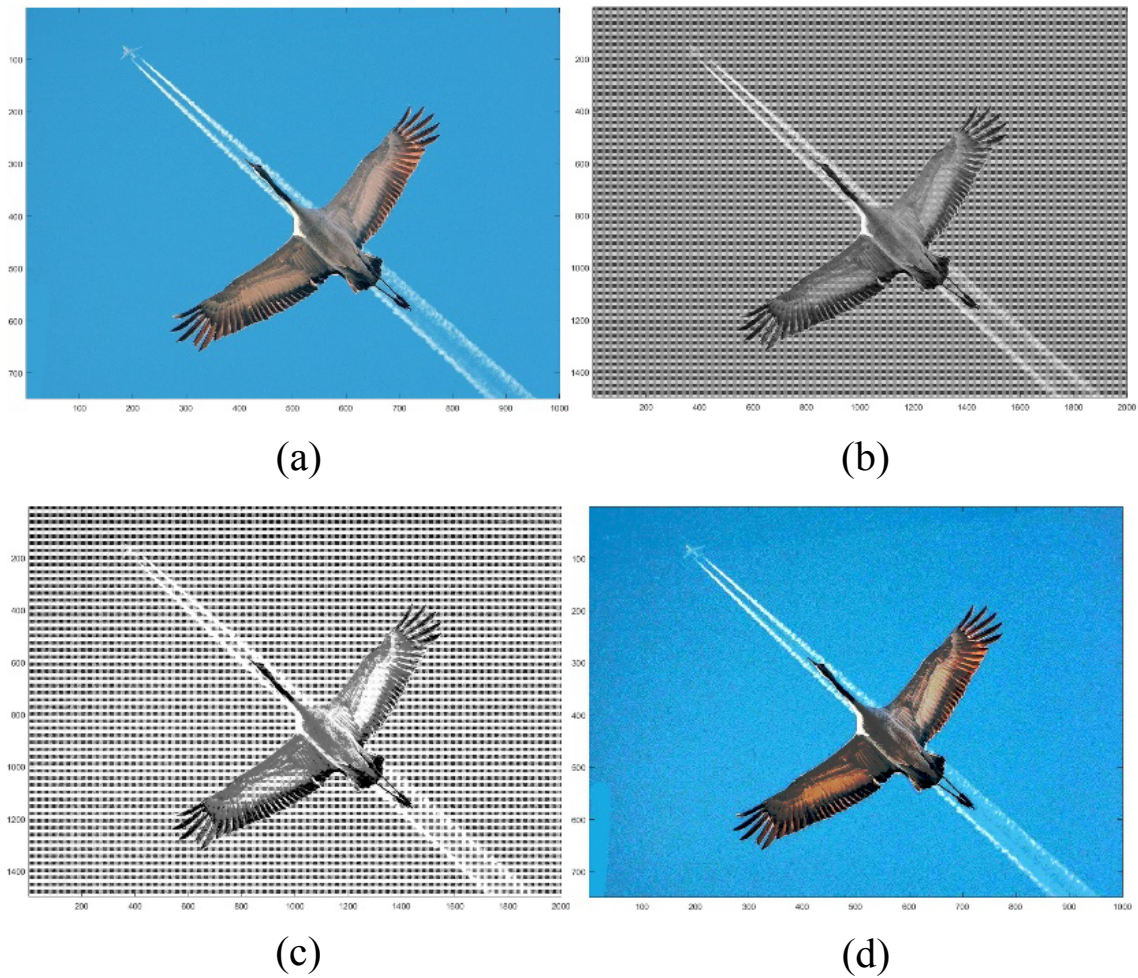

**Figure 13:** a) Original Image "crane.jpg"; b) 2-D Grayscale image transformed by 2 × 2 model; c) Histogram equalized 2-D grayscale image in (b); d) RGB color image after converting the 2-D grayscale image in (c) to color image.

Figure 11 shows results of column model transformation (Figure 11b) of the original color image (Figure 11a). After transforming to the column model 2-D grayscale image, the histogram equalization is calculated on the image; the enhanced grayscale image is as shown in Figure 11c and the corresponding color image in Figure 11d.

Figure 12 shows the row transformed 2-D grayscale image (Figure 12b) of the orignal image (Figure 12a). The image processing is done on the transformed grayscale image and the histogram equalized grayscale image is shown in Figure 12c and the corresponding color image in Figure 12c is shown in Figure 12d.

Table 3 shows that the CEME value of the histogram equalized color image is slightly higher than the CEME of the original image. The CEME values of the 2 × 2 column and row model are exactly the same value. But the CEME value of the 2 × 3 model is slightly different but close to the other CEME values. EME value of the histogram equalized grayscale image is slightly higher than the transformed 2-D grayscale image of the original image. But the EME value of the histogram equalized grayscale image is different for the transformation done by different models.

In Figure 13, the original image "crane.jpg" (Figure 13a) is first transformed to 2-D grayscale image by 2 × 2 model (Figure 13b). Then, the histogram equalization is done on the transformed grayscale image and is shown in Figure 13c. Figure 13d shows the color image of the histogram equalized image in Figure 13c.

Figure 14 shows the image after transforming the im-age "crane.jpg" in (Figure 14a) by 2 × 3 model (Figure 14b). The enhancement algorithm of histogram equalization is done on the transformed grayscale image and the enhanced grayscale image is shown in Figure 14c. This image converted back to the color image and is shown in Figure 14d.

Figure 15 and Figure 16 illustrate respectively the image results after transforming the original color image by column transformation model and row transformation model. The transformed image of the original image in Figure 15a and Figure 16a is shown respectively in Figure 15b and Figure 16b. The histogram equalized gray-scale image is shown in Figure 15c and Figure 16c. The enhancement effects by histogram equalization on color image is shown in Figure 15d and Figure 16d.

Table 3: EME and CEME values of histogram equalization of "flower.tif".

| Image "Flower.tif" | EME | CEME |
|---|---|---|
| Original Image | - | 28.7506 |
| 2-D Grayscale Image (2 × 2 model) | 22.1549 | - |
| 2-D Grayscale Image (2 × 2 model) - After histogram equalization | 23.7389 | - |
| RGB Color image - After histogram equalization of 2-D Grayscale Image (2 × 2 model) | - | 30.0109 |
| 2-D Grayscale Image (2 × 3 model) | 23.6574 | |
| 2-D Grayscale Image (2 × 3 model) - After histogram equalization | 24.6945 | |
| RGB Color image - After histogram equalization of 2-D Grayscale Image (2 × 3 model) | | 29.4427 |
| 2-D Grayscale Image (column model) | 23.0670 | |
| 2-D Grayscale Image (column model) - After histogram equalization | 24.7007 | |
| RGB Color image - After histogram equalization of 2-D Grayscale Image (column model) | | 30.0109 |
| 2-D Grayscale Image (row model) | 23.4101 | |
| 2-D Grayscale Image (row model) - After histogram equalization | 25.0441 | |
| RGB Color image - After histogram equalization of 2-D Grayscale Image (row model) | | 30.0109 |

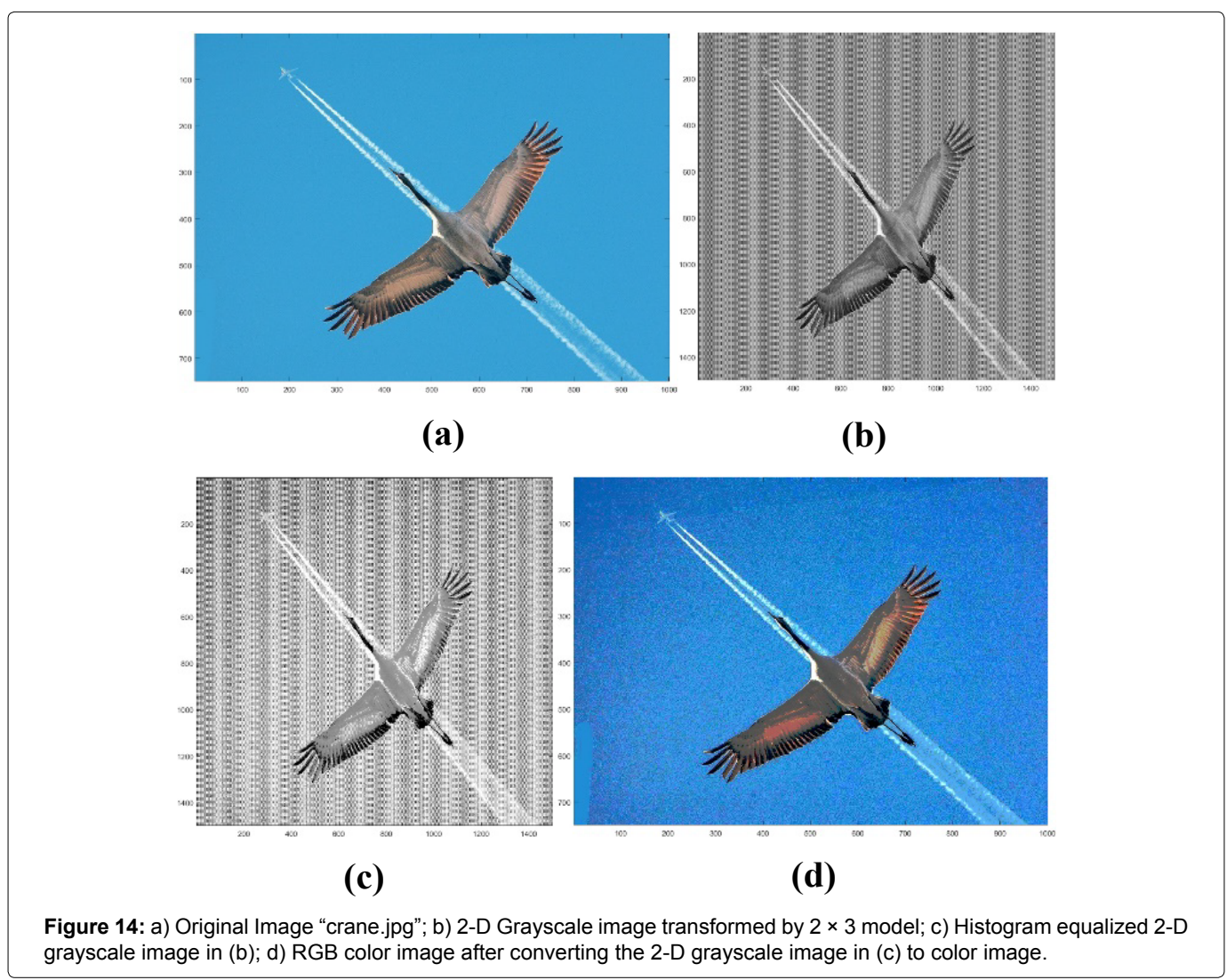

**Figure 14:** a) Original Image "crane.jpg"; b) 2-D Grayscale image transformed by 2 × 3 model; c) Histogram equalized 2-D grayscale image in (b); d) RGB color image after converting the 2-D grayscale image in (c) to color image.

Table 4 shows that the CEME value of enhanced color image is much higher than the CEME value of the original image. In this case also, CEME values of enhanced color image by 2 × 2, column and row model are exactly the same, and the CEME value of 2 × 3 model is slightly different but close to the rest of CEME values. The EME value of the enhanced grayscale image is also higher than for the corresponding grayscale image of the color image. The EME values are showing nearby values for the images transformed by the different transformation model.

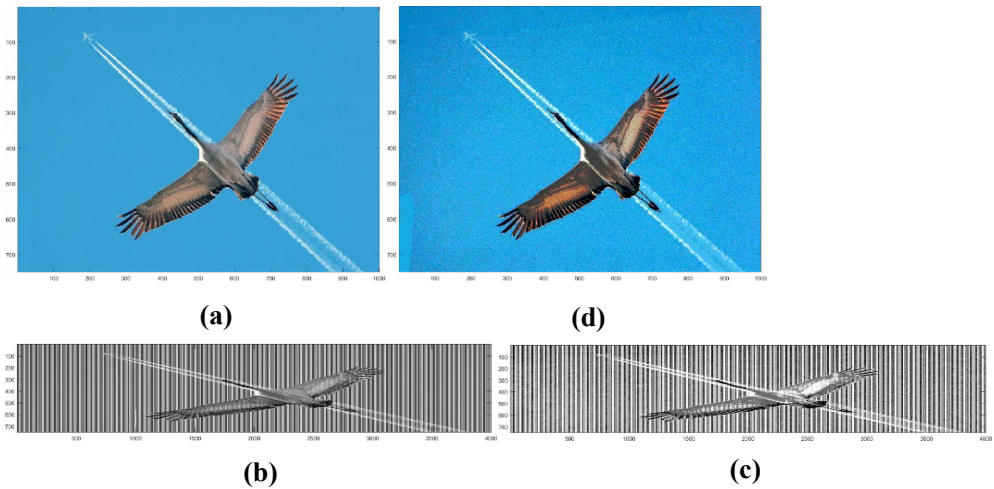

**Figure 15:** a) Original Image "crane.jpg"; b) 2-D Grayscale image transformed by column model; c) Histogram equalized 2-D grayscale image in (b); d) RGB color image after converting the 2-D grayscale image in (c) to color image.

**Table 4:** EME and CEME values of histogram equalization of "crane.jpg".

| Image "Crane.jpg" | EME | CEME |
|---|---|---|
| Original Image | - | 28.7627 |
| 2-D Grayscale Image (2 × 2 model) | 27.4693 | - |
| 2-D Grayscale Image (2 × 2 model) - After histogram equalization | 50.4364 | - |
| RGB Color image - After histogram equalization of 2-D Grayscale Image (2 × 2 model) | - | 44.7457 |
| 2-D Grayscale Image (2 × 3 model) | 27.8003 | |
| 2-D Grayscale Image (2 × 3 model) - After histogram equalization | 49.9529 | |
| RGB Color image - After histogram equalization of 2-D Grayscale Image (2 × 3 model) | | 48.8374 |
| 2-D Grayscale Image (column model) | 27.6265 | |
| 2-D Grayscale Image (column model) - After histogram equalization | 50.4988 | |
| RGB Color image - After histogram equalization of 2-D Grayscale Image (column model) | | 44.7457 |
| 2-D Grayscale Image (row model) | 27.6164 | |
| 2-D Grayscale Image (row model) - After histogram equalization | 50.6269 | |
| RGB Color image - After histogram equalization of 2-D Grayscale Image (row model) | | 44.7457 |

**Table 5:** CEME values of original and enhanced image of "tree.tiff".

| Image "Tree.tiff" | CEME |
|---|---|
| Original Image | 25.9192 |
| Alpha-rooting Method (alpha = 0.98) done on 2-D Grayscale Image (2 × 3 model) of RGB model of original image | 34.1776 |
| Alpha-rooting Method (alpha = 0.97) done on 2-D Grayscale Image (2 × 3 model) of XYZ model of original image | 35.3251 |
| Alpha-rooting Method (alpha = 0.99) done on 2-D Grayscale Image (2 × 3 model) of CMY model of original image | 23.4129 |
| Alpha-rooting Method (alpha = 0.96) is done on Y component of YUV Color model of original image | 34.6687 |
| Histogram Equalization is done on 2-D Grayscale Image (2 × 3 model) of RGB Color model of original image | 26.7260 |
| Histogram Equalization is done on 2-D Grayscale Image (2 × 3 model) of XYZ Color model of original image | 15.3764 |
| Histogram Equalization is done on 2-D Grayscale Image (2 × 3 model) of CMY Color model of original image | 26.7260 |
| Histogram Equalization is done on Y component of YUV Color model of original image | 13.2658 |

We also did a comparative study of enhancement by the proposed method for different color models. Figure 17 show enhanced image results of the proposed algorithm when applied to the RGB, XYZ and CMY color models of the original image "tree.tiff" for frequency domain enhancement technique alpha-rooting method. While Figure 18 show the image results for spatial domain enhancement techniques histogram equalization by the proposed method for image "tree.tiff" in the color models RGB, XYZ, and CMY. The transformation model used in Figure 17a, Figure 17b, Figure 17c, Figure 17d, Figure 18a, Figure 18b, Figure 18c and Figure 18d is the 2 × 3 model. Figure 17e and Figure 18e show respectively image results of alpha-rooting method and histogram equalization on the color model YUV. In YUV model, the Y component is enhanced by the enhancement technique. Table 5 shows the CEME values of the original and enhanced image of "tree.tiff" for both alpha-rooting method and histogram equalization for the color models RGB, XYZ, CMY, and YUV.

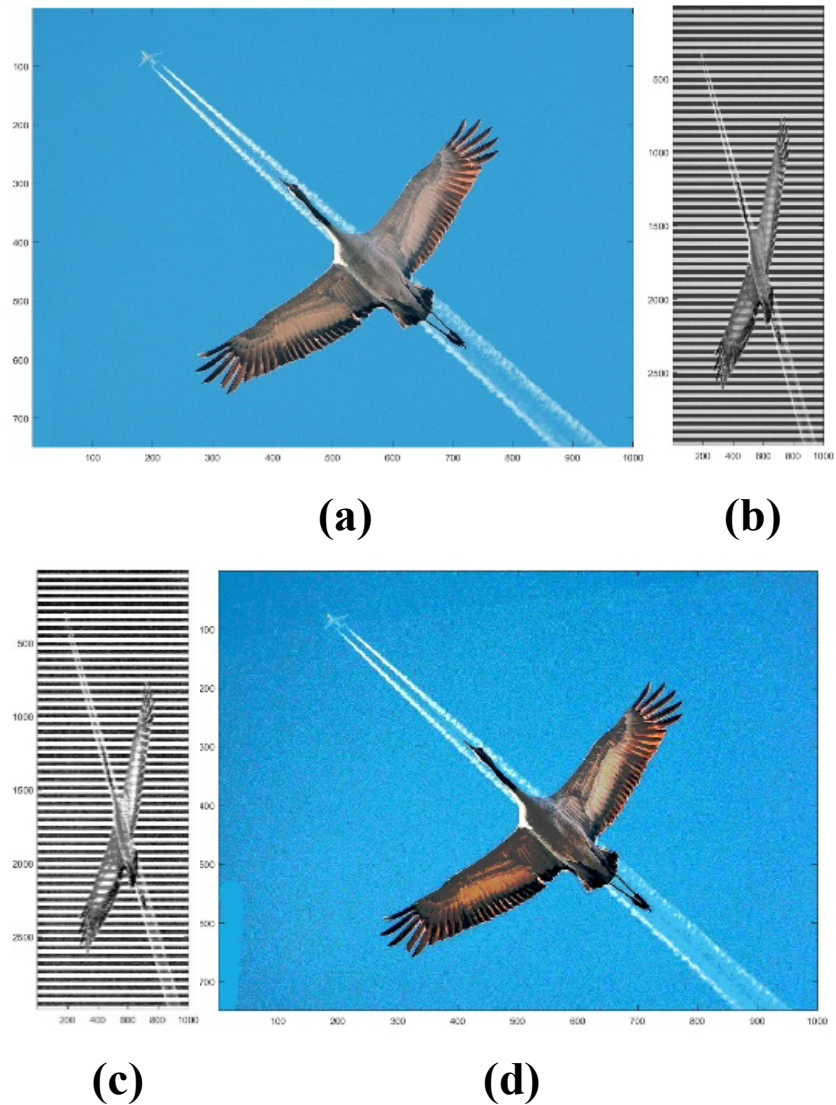

**Figure 16:** a) Original Image "crane.jpg"; b) 2-D Grayscale image transformed by row model; c) Histogram equalized 2-D grayscale image in (b); d) RGB color image after converting the 2-D grayscale image in (c) to color image.

## Conclusions

The novel approach to of color image enhancement shows good enhancement results for enhance-ment algorithms in both frequency domain as in the alpha-rooting method, and in the spatial domain, as in the histogram equalization. The CEME metric shows that enhancement by all different transformations gives better CEME than the original image. It is also seen that the CEME value of the enhanced image gives almost nearby range values for all different models. In spatial domain enhancement by histogram equalization, it is seen that the CEME value of the enhanced image is exactly the same value for color image transformed by the proposed 2 × 2, column and row models and slightly different from that of 2 × 3 model. This is because in the 2 × 3 model, we do not consider the luminous part $I(n, m)$ separately. But in all other models, the $I(n, m)$ is also included in experimental methods.

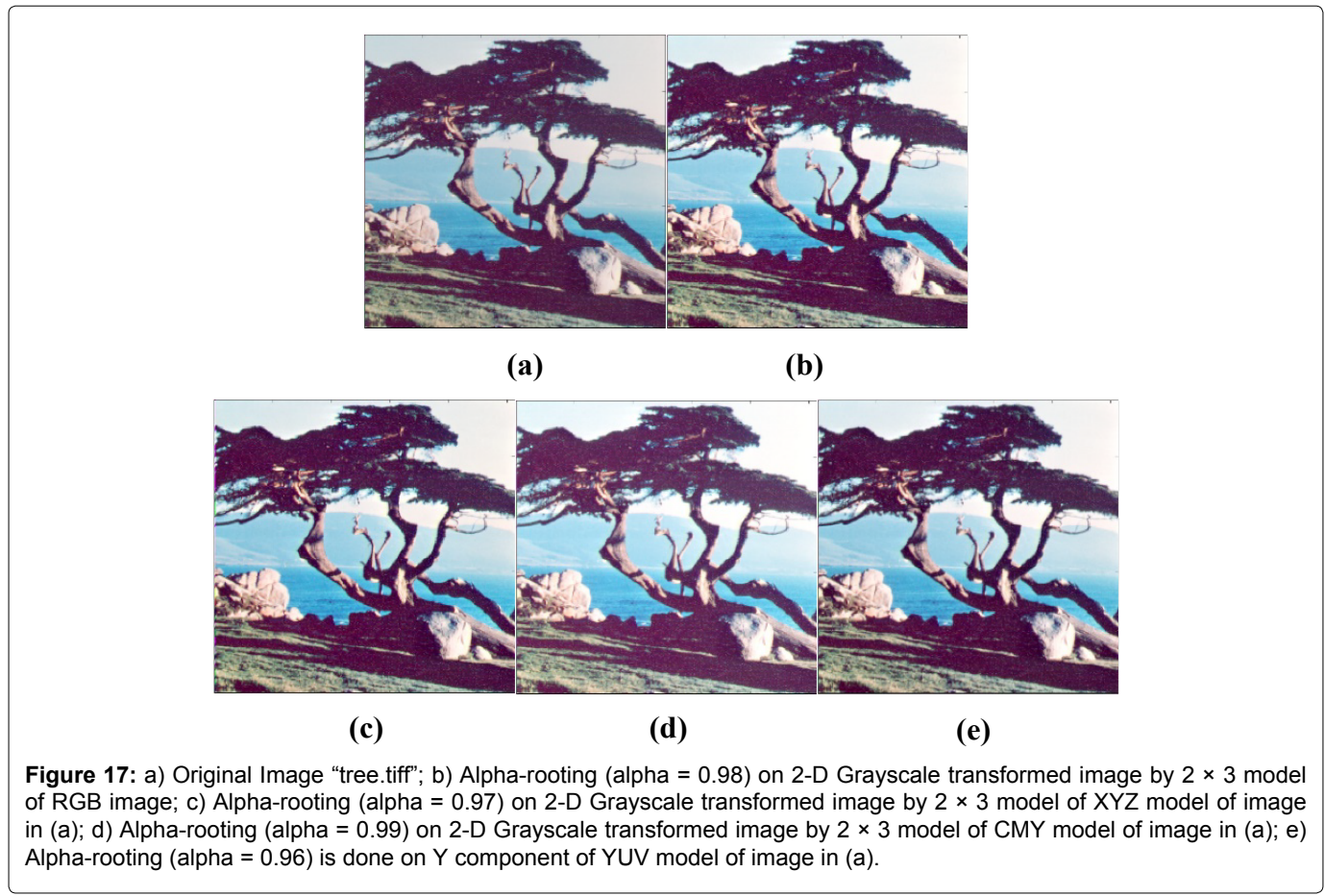

**Figure 17:** a) Original Image "tree.tiff"; b) Alpha-rooting (alpha = 0.98) on 2-D Grayscale transformed image by 2 × 3 model of RGB image; c) Alpha-rooting (alpha = 0.97) on 2-D Grayscale transformed image by 2 × 3 model of XYZ model of image in (a); d) Alpha-rooting (alpha = 0.99) on 2-D Grayscale transformed image by 2 × 3 model of CMY model of image in (a); e) Alpha-rooting (alpha = 0.96) is done on Y component of YUV model of image in (a).

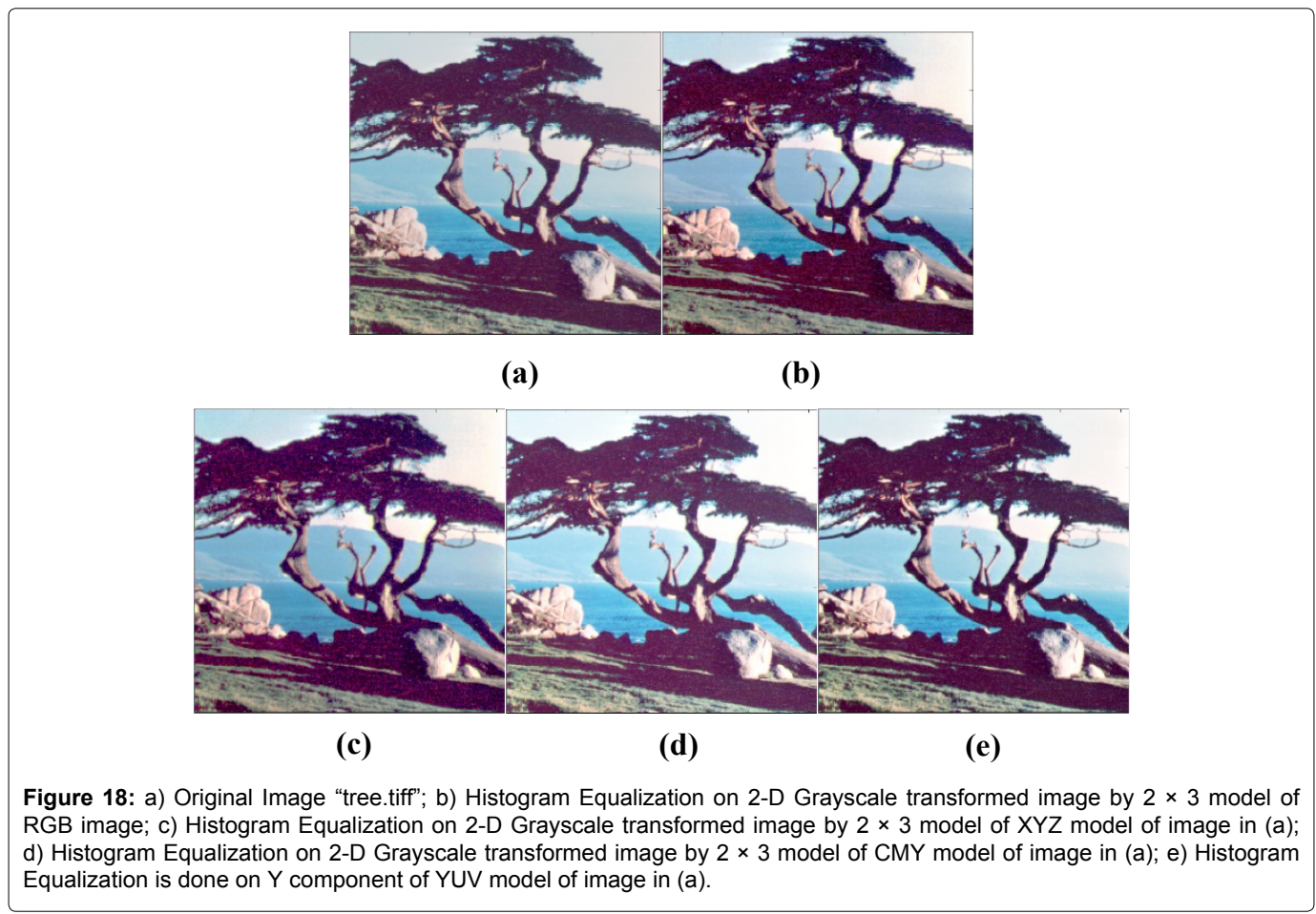

**Figure 18:** a) Original Image "tree.tiff"; b) Histogram Equalization on 2-D Grayscale transformed image by 2 × 3 model of RGB image; c) Histogram Equalization on 2-D Grayscale transformed image by 2 × 3 model of XYZ model of image in (a); d) Histogram Equalization on 2-D Grayscale transformed image by 2 × 3 model of CMY model of image in (a); e) Histogram Equalization is done on Y component of YUV model of image in (a).